# An Effective Approach to Scramble Multiple Diagnostic Imageries Using Chaos-Based Cryptography

Chandra Sekhar Sanaboina

Assistant Professor, Department of Computer Science and Engineering,
University College of Engineering Kakinada(A),
Jawaharlal Nehru Technological University Kakinada, 533003, India.
Email: chandrasekhar.s@jntucek.ac.in

Tejaswini Yadla

Department of Computer Science and Engineering,
Cyber Security, PG Scholar,
University College of Engineering Kakinada(A),
Jawaharlal Nehru Technological University Kakinada, 533003, AP, India.
Email: tejaswiniyadla99@gmail.com

Medical image encryption could aid in preserving patient privacy. In this article, we provide a chaotic system-based medical picture encryption method. The diffusion and permutation architecture was used. The permutation based on plain image and chaotic keys is offered to shuffle the plain picture's pixels to other rows and columns, successfully weakening the strong connections between neighboring pixels. Diffusion is suggested to spread small changes of plain images to all of the pixels in cipher images to enhance the encryption effect. We analyze the chaotic behavior of the proposed system using various techniques and tests such as bifurcation plots, Lyapunov exponents, MSE, PSNR tests, and histogram analysis.

## 1. INTRODUCTION

More and more individuals are using technology today for communication, data storage, and transport, including computers, smartphones, and a range of other devices. As a result, there is an increase in the quantity of users as well as the quantity of unauthorized users trying to gain unfair access to data. As a result, the question of data security is brought up. The solution to this problem is to communicate or store data in an encrypted format. The data is encrypted, so unauthorized users cannot decipher it. Data is protected while being transported and stored thanks to the field of information security known as cryptography.

The two elements of each encryption and decryption method are the algorithm and the key used for encryption and decryption. But what makes cryptography secure is the key that is used for both encryption and decoding. There are two distinct types of cryptographic procedures in symmetric key cryptography, where one key is used for both encryption and decryption. Asymmetric key cryptography encrypts and decrypts data using two different keys. In comparison to the asymmetric key algorithm, the symmetric key algorithm is significantly faster, simpler to build, and consumes fewer processing resources.

### 1.1. Telemedicine

Medical imaging technology is a significant component of the diagnostic process used in digital health to produce two- or three-dimensional images. In the modern healthcare system, it is frequently a problem for organizations that various doctors cannot access each other's patient information. To conserve resources and produce chronologically accurate medical reports for each patient, it is especially important to minimize multiple acquisitions of practically identical medical picture data as well as the loss of such prior data.

A distributed database architecture that would enable all clinicians to electronically access all current medical information for their patients, including all medical imaging data amassed over many years, would be one answer to these issues.

The urgency with which patient-related medical imaging data must be protected and kept confidential while being held in databases and sent over any type of network is amply illustrated by the facts. Patient personal information may be lost if these photos are used improperly.

### 1.2. Basic Cryptography Principles:

Cryptography is the foundation of image security. For applications in image security, several foundational ideas from cryptography are employed as basic components (primitives). An introduction to cryptography is given in order to better grasp the challenges surrounding picture security. (See Fig 1).

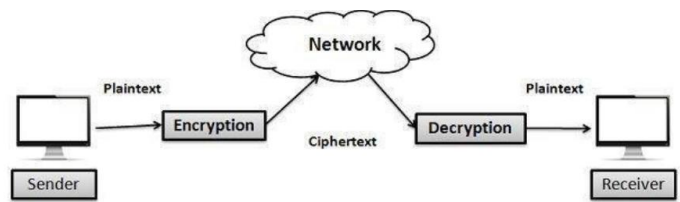

Figure 1: Cryptography

### 1.3. Characteristics of Image Encryption Methods

Image data is very redundant and strongly linked between pixels, unlike text communications. They are often enormous, too. These factors add together to make standard encryption technologies sluggish and challenging to use. Real-time processing, accuracy assurance, picture format consistency, and data reduction for transmission are just a few of the demands that imaging applications could have. Real-time imaging techniques have faced substantial difficulties in meeting these needs while still maintaining strong security and quality standards. Look into picture encryption. First, we must examine how text and picture data are implemented differently in terms of encryption. Encrypting text data and picture data has several differences.

1. 1. The encrypted picture can, however, be lossily decrypted to return to the original plain image.
2. A phrase is written data. Block ciphers or stream ciphers can be used for direct encryption. The data must be transformed to a 1D array before employing several conventional encryption techniques in order to secure a 2D array of data saved using word processing algorithms.
3. 3. Since picture storage is quite huge, it could not be efficient to directly encrypt or decode photos. One of the greatest methods to cut down on both storage space and transmission time is to encrypt and decrypt data that is only utilized for picture compression.

## 2. RELATED WORK

As in previous years, there are more applications for digital data transfer and picture transmission, therefore ensuring secure image transmission is crucial. To give picture applications security, image encryption is used. It provides an overview of the several image encryption methods that are currently in use. Additionally, it outlines a few traits of effective picture encoding methods. outlines the most well-known algorithms and research studies about different image encryption methods.

M. K. Hasan *et al.*(Hasan et al., 2021) reviewed the many encryption methods now in use, taking into account execution time, memory demand, and encryption quality.

B. Vaseghi *et al.*(Vaseghi et al., 2021) A chaotic cryptosystem is offered to increase the security of medical picture transmission and storage, and it uses synchronized chaotic maps as private key generators. Using simulation and analytical investigation, the usability and effectiveness of the suggested synchronization strategy are evaluated.

Ravichandran *et al.*(Ravichandran et al., 2017) proposed a hybrid cryptographic method utilizing chaotic maps and deoxyribose that may be adjusted for both partial and complete medical picture encryption. The very random keys for encrypting the color digital imaging and communications in medicine pictures are produced by the suggested approach by combining several chaotic maps in a single step.

P. Deshmukh(Deshmukh, 2016) To prevent unwanted access to the picture data, the AES algorithm for image encryption and decryption has been applied. One of the finest encryption and decryption standards on the market is the effective use of the symmetric key AES algorithm. An AES method is synthesized and simulated for image encryption and decryption with the aid of MATLAB code.

A. A. Abd El-Latif *et al.* (Abd El-Latif et al., 2017) a fresh strategy for effective quantum image encryption of medical material. The suggested approach makes use of a chaotic map and grey coding. Quantum grey coding jumbles the quantum picture. Then, a quantum XOR operation based on a key generator controlled by the logistic-sine map is used to encrypt the jumbled quantum picture. The proposed encryption and decryption algorithm's circuits are designed using a NEQR quantum image representation.

K. Shankar *et al.*(Shankar et al., 2018) A security model was put up to address the issue of 1-D chaotic cryptosystems' limited key space and feeble security. This study used chaotic logistic and tent maps to analyze extremely secure medical pictures with a few subkeys where a few subkeys are provided. The security was examined using the chaotic (C-function) process, including diffusion and confusion.

S. T. Kamal *et al.*(Kamal et al., 2021) introduce a novel encryption method that uses picture blocks and chaos to encode medical images in both grayscale and color. A brand-new picture-splitting method built on image blocks was presented. Then, a zigzag pattern, rotation, and a randomized permutation are used to jumble the picture blocks. Then a key to decode the jumbled image is generated via a chaotic logistic map.

C. H. Lin *et al.*(Lin et al., 2021) proposed intelligent symmetrical cryptography with only a chaotic map and a key generator based on quantum mechanics for encrypting and decrypting medical images. It also comprises the creation of random cipher codes, evaluation of decrypted images, and training of encryptors and decryptors based on grey relational analysis.

Y. Ding *et al.*(Ding et al., 2021) presented medical picture encryption, and a decryption technique that is a deep learning based is suggested. To encrypt and decode the medical image, the Cycle-Generative Adversarial Network is employed as the main educational network.

To summarise, a variety of methods and tactics were created and implemented for effective data security, each of which has pros and cons. Time complexity, computing cost, attack robustness, dynamical behavior, and security are the most often used assessment and selection criteria. A better set of encryption rules for the security of clinical data is provided by the research reported in this paper and might be employed in a trustworthy e-healthcare device. The major goal is to create a simple data encryption method with low key sensitivity, minimum residual consistency, and high-quality data retrieved using chaotic maps.

The rest of this manuscript is divided into the following sections. Information about suggested chaotic maps' characteristics and behaviors is provided in Section III. This is proven through the use of the proposed photo encryption/decryption device's information and comprehensive description. Simulation experiments and thorough quantitative and qualitative analysis are found in Section IV. The conclusion is located in Section V.

## 3. PROPOSED METHOD

### 3.1. Chaotic Maps

*Chaos Theory:*

Non-linear dynamical systems that show sensitivity to beginning circumstances are the subject of the mathematical field known as chaos theory. Nature contains chaotic situations. The weather is a good and typical illustration of this. Since its introduction, chaos theory has found extensive use in a variety of fields, including computer science, ethnography, economics, and meteorology. Sensitivity to beginning circumstances, periodic orbit density, and topological mixing are characteristics of chaotic systems. Chaotic systems are appealing for cryptography because of their qualities like determinism and sensitivity.

In modern cryptography-associated research work a growing number of people are using chaotic algorithms due to their intrinsic qualities or traits. In this study, we utilized pipelines for data encryption and/or chaotic maps for reliable medical imaging. The framework that is being proposed is a discrete-time, nonlinear strategy with distinctive dynamical chaotic behavior. Compared to probability distribution-based algorithms, searches can be carried out more quickly because of the inherent chaos in unrepeatability and ergodicity properties. Two unique maps are utilized and researched for medical image encryption among other potential maps.

Any function that displays chaotic behavior is, by definition, a chaotic map. The mathematically controlled chaotic behavior of the suggested maps defines a given point that maps ($x_n$, $y_n$) to a new location ($x_{n+1}$, $y_{n+1}$):

$$y_{n+1} = y_n - r * \tanh x_n \quad (1)$$
$$x_{n+1} = \sin x_n + \cos y_n$$

$$y_{n+1} = b * x_n^2 \quad (2)$$
$$x_{n+1} = x_n + y_n^2 - a * r$$

Starting with one of the models provided by Eqs. (1) or (2), maps are generated by defining the starting system parameters (n, the upper and lower limits, the number of dimensions, and the fitness function). The map's initial locations $x_0$ and $y_0$ are then generated at random. Finally, $x_{n+1}$ and $y_{n+1}$ are updated via an iterative procedure that is repeated n times.

Additionally, non-smooth trajectories and interrupted motion are features of maps' dynamic characteristics. The suggested method converts the original two-dimensional picture into a sequence of bytes that are then encoded using an iterative approach.

Eqs. (1) or (2) use the proposed chaotic functions (2). The technique is backed by a few logical procedures that improve security and accelerate encryption. One is used to modify the pixel locations, while the other is used to modify the pixel density value image-splitting operations, including the mergers and acquisitions necessary for encryption.

### 3.2. Suggested Encryption and Decryption Structure:

We provide a brand-new encryption and decryption method for images based on chaos. The heart of the suggested approach is an image-splitting technique, in which the pixels of an input picture that has to be encrypted is split into various parts ($I_s$, s ≥ 2). We select s = 2 so that each half of the image is encoded using one of the two innovative keymaps that we introduce. Generally speaking, any quantity of splits and various maps can be utilized to encode various regions of an image.

An appropriate set of Is will, at least in part, aid in the regeneration of *I*. This method is based on the premise that individual picture sections or splits do not communicate any useful information. The recovered image may have low quality due to loss of color and contrast while being a successful strategy. A new, quick method that uses innovative chaotic keys and has the benefit of having a lower error rate throughout the encryption and decryption stages is proposed to partially overcome this constraint. This method can preserve the quality of the restored image. The suggested techniques additionally maintain attack resistance while having low execution time and bandwidth needs. Two phases of permutation (also known as confusion) and diffusion procedures are used to fully encrypt an image. Both processes are created so that chaotic states and plain image data are employed, respectively, to shift pixel placements and substitute pixel values, resulting in a cipher image that resembles noise.

*Encryption Process*

Our chaotic medical picture encryption technique's flow diagram is shown in Fig. 2. A 1D matrix is created from a specified input plain image (size is D×H×W) by adding the picture pixel values collectively columns-wise. Here, H and W denote the height and width of the image, respectively. while D stands for the depth of the image, with D being equal to 1 for grayscale photos and color photos 3.

To process, the produced 1D matrix is partitioned into two sections, each measuring (D×H×W)/2, for (P1 and P2). The two suggested maps are used in the specified coding scheme to encrypt separated sections. The suggested maps are provided by (1) and (2), where r, a, and b are the external parameters that control the chaotic behavior of the map and are utilized as keys in the encryption scheme. For encoding individual picture splits, several maps may be utilized alone or in combination. The chaotic key sequences and XOR are used to encrypt the divided images.

*Permutation Process*

Chaotic maps are used to produce turbulent groups for the permutation procedure. The chaotic sequences are utilized to build a network of records, which is then applied to rearrange the original image to produce the thinned image. It is challenging to properly identify the image because when the ordered image is acquired, the relationship between the neighboring pixels is complete. As a result, utilizing the dispersion strategy improved security. The preceding cipher pixel value and the associated keystream element both control pixel changes. Because of this, even the smallest changes to one pixel might have an impact on all subsequent pixels. The 1D splitter matrix substitutes are XORed with the key sequence. The permutation technique is repeated three more times with a random key sequence.

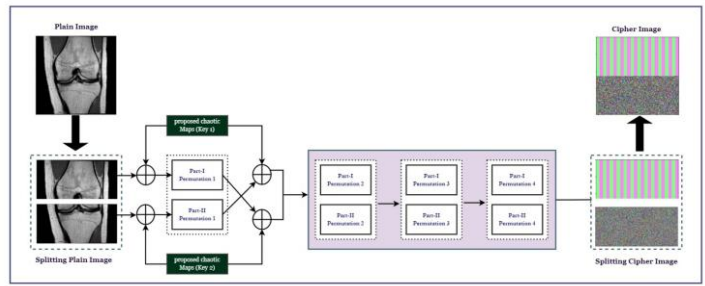

Figure 2: Flow Diagram Illustrating the Encryption Process for the Proposed Scheme

---

Algorithm 1: Proposed Algorithm for Encryption Process

---

1. The output is the encrypted noisy picture C, input is the original image P (size H × W).
2. Transform P into a 1D array of pixels and divide it into vectors P1 and P2 which are each H × W/2 in size.
3. Equations (1) and (2) are put into practice n times (n is the length of the array) to produce the chaotic sequence.
4. Modify and change while keeping in mind that each component of the grayscale falls inside the range [0, 255], using the suggested chaotic maps in place of Step 3 and the associated condition (3)

$$P(t) = \text{mod}(\text{round}(10^{12} P(t)), 256), \text{ where } 1 \leq t \leq H \times W/2. \quad (3)$$

5. Using the XOR operation (4), generate diffused vector.

$$D_i = P_i \oplus K_i \ (\oplus \text{ is the bit-wise XOR}) \quad (4)$$

6. Permute the sequence of $P_1$ with the 1$^{st}$ vector of the 1$^{st}$ proposed map and $P_2$ with the 1$^{st}$ vector of the 2$^{nd}$ proposed map (5).

$$p_i = D_i(K_i) \quad (5)$$

7. Swap the splits and again generate diffused vector by (4).
8. Again Re-Permute (3 times) the sequence of $P_1$ with the 2$^{nd}$ vector of the 1$^{st}$ proposed map and $P_2$ with the 2$^{nd}$ vector of the 2$^{nd}$ proposed map (5).
9. Join the two encrypted parts to get the cipher picture, and create the final matrix with (6).

$$C = \text{reshape}(p_i, H, W) \quad (6)$$

*Decryption Process*

The process of decryption is essentially the inverse of the process of the encryption stage. Fig. 3, shows a schematic representation of our decryption stage. It is possible to produce chaotic record sets and disordered vectors produced by the encryption process using the same enigmatic keys. Inverse diffusion, confusion, and combination are used to begin the decryption process.

The encoded image C is initially transformed into a 2D matrix with the dimensions H W, and a diffuse inverted vector is produced in the middle. Additionally, we obtain the common alternating vector and split it into two pieces ($P1_p$, $P2_p$). Finally, using the disorderly recording arrangement and a mixture of methods, each partial area's inverse level ($P_1$, $P_2$) was obtained.

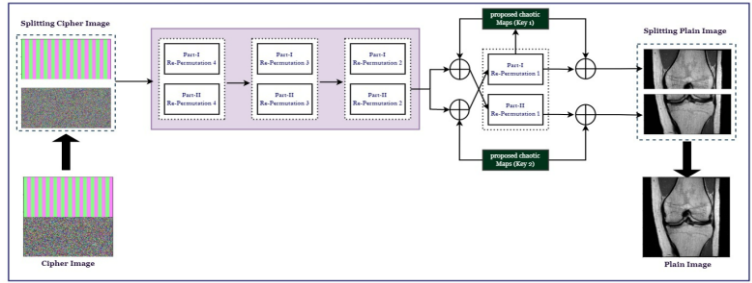

Figure 3: Flow Diagram Illustrating the Decryption Process for the Proposed Scheme

Algorithm 2: Proposed Algorithm for Decryption Process

1. The input is the encrypted noisy picture C and the original image P is the output
2. Transform C into a 1D array of pixels and divide it into vectors C1 and C2 which are each H × W/2 in size
3. Inverse permute the vector 3 times with (7)

$$D_i(K_i) = c_i \quad (7)$$

4. Generate the de-shuffled vector using (8)

$$C = D \oplus K \text{ (bit-wise XOR operation } \oplus) \quad (8)$$

5. Generate an Inverse permuted vector where C=K (index) by XOR the second chaotic sequence with the 1$^{st}$ proposed map to scramble $C_1$, and the second chaotic sequence with the 2$^{nd}$ proposed map to scramble $C_2$.
6. Swap the splits and again, create an inverse permutation of the sequences C1 and C2 using the first vectors from the first and second proposed maps, respectively (7).
7. Re-generate the de-shuffled vector using (8)
8. Create C1 and C2 using the chaotic index sequence and vector component reshaping.
9. Combine resultant halves, $C_1$, and $C_2$, to get P.

4. EXPERIMENTAL RESULTS

New maps have been used to construct the suggested encryption and decryption procedure, and it has been assessed using various MRI knee images. Fig. 4 shows the outcomes of encryption and decryption. An Intel(R) machine with a Core (TM) i5-24004 CPU, 8 GB of RAM, and a 512 GB SSD was used to actualize the simulation platform. The Google Colab Python application was utilized for all simulation studies. Not to mention that each simulation test has been performed many times.

Fig 4 illustrates the encryption and decryption procedures step-by-step using a knee MRI scan. The figure shows different application processing steps of the proposed scheme. Controlling parameter values must be chosen since different combinations of these parameters provide disorganized results when using the proposed maps. We used r = 17.0 and 2.35 for Eqs. (1) and (2), respectively, in our simulation studies. The places where chaos behavior can be seen in the bifurcation and Lyapunov exponent diagrams in Fig. 4 were chosen as the locations for the r values for both maps. Other values may be utilized and have comparable or almost the same results, as Fig suggested. After multiple experiments, the best a and b values of Eq have been selected (2). In particular, simulation tests were run with different sets of a and b, [0.1: 0.1: 1], and map trajectories were evaluated. The optimal a and b values a = 0.5, b = 0.3 where the map performs well under chaotic conditions—are then chosen. We choose $x_0$ and $y_0$ to be equal to 0.1 for the initial map values, which can be any arbitrarily chosen numbers. The number of iterations was set at n=1000 to construct a chaotic sequence that performed well at random. Fig 4(a) shows the original medical image (MRI knee scan) which needs to be encrypted and Fig 4(b) is its split parts. Fig 4(c) and 4(d) show the encrypted form of the original medical image (MRI knee scan) and its split parts Fig 4(e) shows the encrypted form of the original medical image. Fig 4

Bifurcation and Lyapunov diagrams are used to examine the chaotic behavior of the suggested maps. The first map's bifurcation and Lyapunov exponent are shown in Figures 5(b) and (c), respectively. The second map's bifurcation and Lyapunov exponent are shown in Figs. 5(e) and 5(f), respectively. One parameter is altered at a time while the others are kept fixed when evaluating bifurcation diagrams. The bifurcation diagrams are explored for parameters *a* and *b*. The bifurcation diagram of the chaotic system with *r* as a varying parameter is shown in Fig 5 where a=0.1 and b=0.1 parameters are kept fixed.

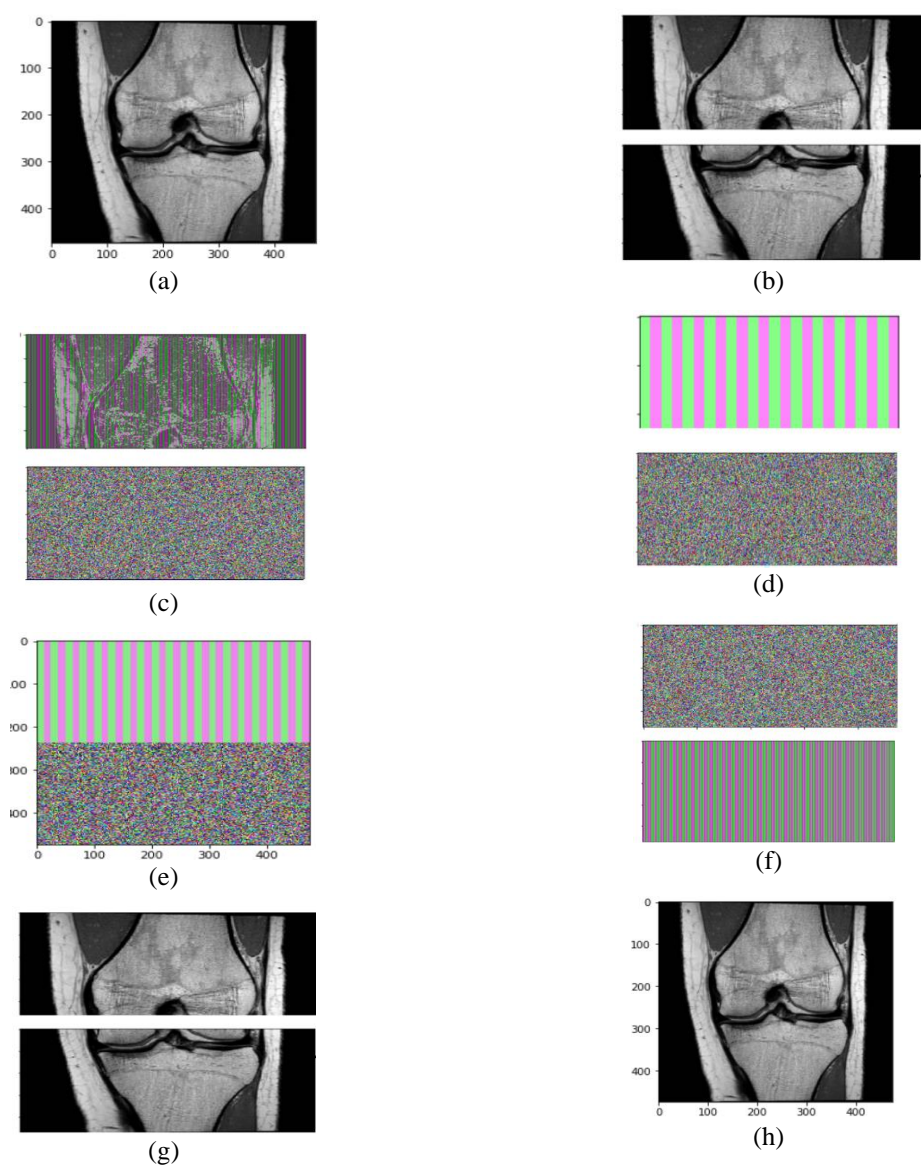

Figure 4: The simulation steps for encryption (a→h) and decryption (a←h) process: (a) original MRI knee image; (b) split images; (c) permutation for each half of split images; (d) Re permutation for the two halves; (e) Combination and final diffusion encrypted image; (f) Inverse permuted image splits; (g) Re inverse permuted splits; (h) Combination and final diffusion decrypted image

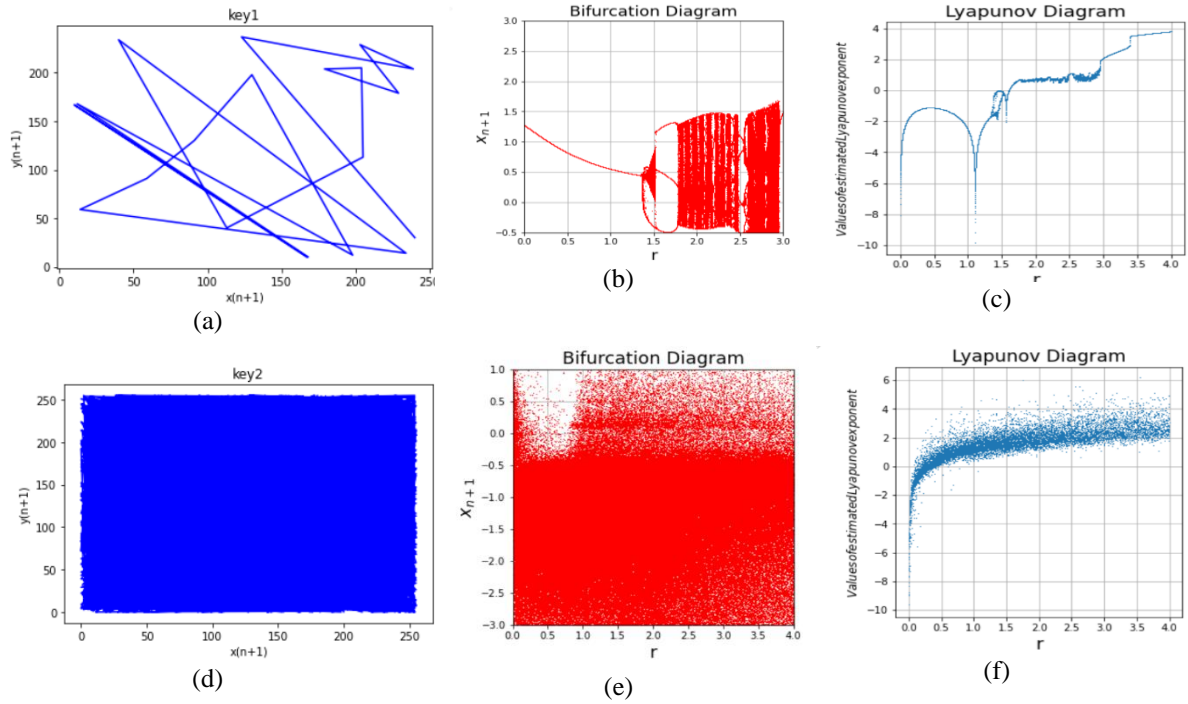

Figure 5: Shows the suggested maps' 2D (x, y) phase plots together with the corresponding bifurcation (b, e) and Lyapunov exponent (c, f) diagrams. Eq. (1) with r = 17.0 produces the map in (a), while Eq. (2) with a = 0.5, b = 0.3, and r = 2.35 produces the map in (d). With $x_0 = y_0 = 0.1$, simulated maps are produced.

### 4.1. Statistical and Qualitative Analyses

The well-known technique of histogram analysis may be used to qualitatively assess a particular system in picture encryption.

The image histogram analysis shown in Fig 6 is one of the simplest ways to demonstrate the quality of picture encryption. A decent image encryption technique usually transforms a plaintext image into a random, unintelligible form. Thus a good image encryption technique generates a cipher image that has a uniformly distributed intensity histogram. Figure 6 shows the histogram analysis for the selected sample medical images and their encrypted equivalents. While the histograms of the encrypted images are dispersed uniformly over the range of greyscale (Fig. 5 (d, h, l, p)), those of the original photos are clustered around certain grey values, as observed in Fig. 6 (c, g, k, o). This demonstrates how resistant to statistical assaults the encrypted data is.

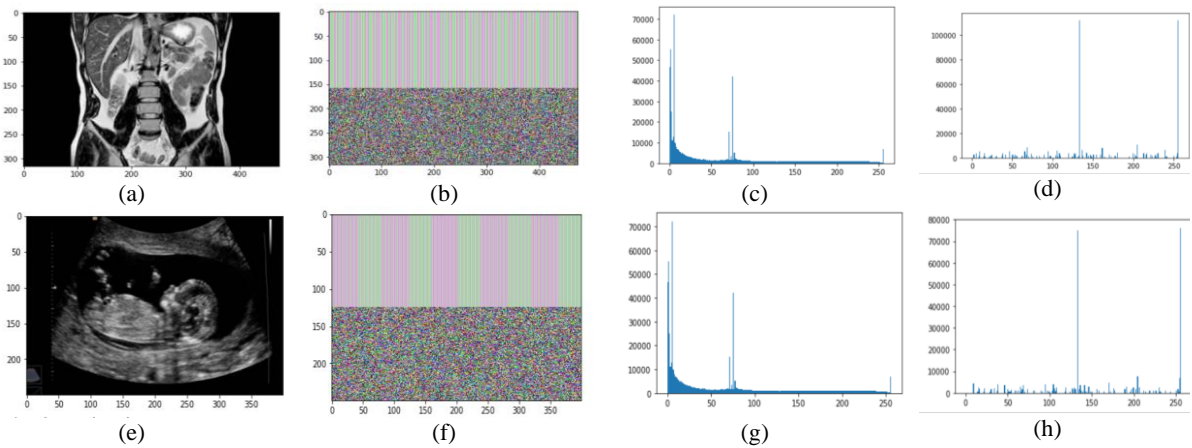

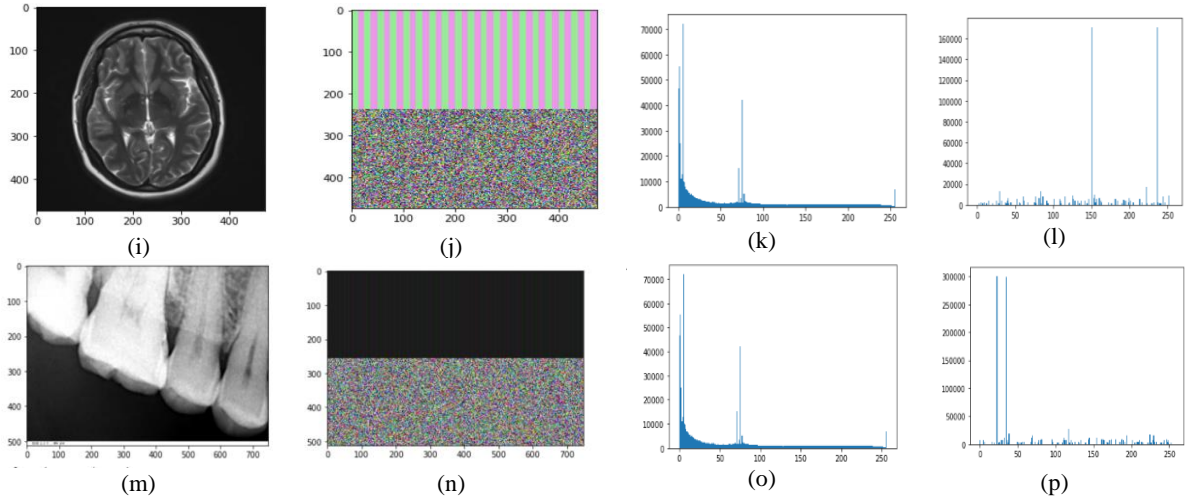

Figure 6: Shows the qualitative outcomes of the suggested approach utilizing different medical photos. - columns from left to right: original images, cipher equivalents, original images histograms, and cipher images histograms. - rows from top to bottom: abdominal CT scan, plain ultrasound image, brain MRI image, and an X-ray of the tooth.

### 4.2. Metrics for the Proposed System's Evaluation

The resilience against differential assaults may be qualitatively evaluated in practice using a variety of indicators. These measurements are used to gauge a cryptosystem's strength.

The robustness against differential attacks can be qualitatively evaluated in practice using a variety of indicators. These measurements are used to determine a cryptosystem's strength.

The Mean Square Error (MSE) is one of those metrics that is used as an evaluation metric. MSE measures the difference between two given images. Let $I_1$ and $I_2$ be original and encrypted images, and m and n are dimensions of an image, The formula for MSE is given as-

$$\text{MSE} = \sum_{m,n}[I_1(i,j) - I_2(i,j)]^2/(m \times n)$$

PSNR (Peak Signal-To-Noise Ratio), in addition to MSE, is frequently used to gauge how much the original and decrypted images have degraded. The PSNR acronym stands for the Peak signal-to-noise ratio. The PSNR formula is provided as-

$$\text{PSNR}= 10 \log_{10}(R^2/\text{MSE}) \text{ where } R=255$$

The result after performing the qualitative analysis of the encrypted and decrypted image is depicted in below Table 1.

Table 1: Metrics used to evaluate the suggested strategy after it was applied to various medical pictures.

| | Test Medical images | | | |
|---|---|---|---|---|
| Metric | Abdomen CT | Ultra-sound | Brain MRI | X-ray of the tooth |
| MSE | 20175.720 | 24481.361 | 18993.713 | 21600.828 |
| PSNR | 5.082 | 4.242 | 5.345 | 4.786 |

### 5. CONCLUSION

This paper presents a novel and safe approach for encrypting medical images that may be used with cloud-based IoH systems (IoHS). Two completely new chaotic maps were produced via the recommended pipeline, and they displayed strong chaotic behaviors, unpredictability, and high sensitivity to beginning seeds. According to dynamic analysis and validation using the bifurcation diagram and Lyapunov exponent, the suggested maps are often hyperchaotic with high complexity and high sensitivity. The proposed pipeline strategy also employs a two-run confusion-diffusion architecture with an extra 3 permutations, and other input parameters besides the plain picture and the secret key, in contrast to encryption methods based on one-time keys. The latter has the benefit of making it possible to control encrypted data values without affecting the secret keys.

Medical imaging scans have been employed among other tests to illustrate this. In conclusion, the proposed framework has the potential to significantly improve both the security of patient data and the aesthetic appeal of medical images. It is important to keep in mind that the suggested pipeline may be used for any multimedia encryption application, even ones unrelated to medicine.

*Future Scope:*

Future research should focus on the superior chaotic behaviors of the hyper-chaos and high-dimensional chaos maps. In order to increase the complexity of the composite discrete chaotic system and produce excellent chaos, we can design composite discrete high-dimensional chaotic systems and composite hyper-chaotic dynamical systems; however, we should take into account the time commitment and practicality when using these systems for image encryption.

## References

Abd El-Latif, A. A., Abd-El-Atty, B., & Talha, M. (2017). Robust Encryption of Quantum Medical Images. *IEEE Access*, *6*, 1073–1081. https://doi.org/10.1109/ACCESS.2017.2777869

Deshmukh, P. (2016). *An image encryption and decryption using AES algorithm*. http://www.ijser.org

Ding, Y., Wu, G., Chen, D., Zhang, N., Gong, L., Cao, M., & Qin, Z. (2021). DeepEDN: A Deep-Learning-Based Image Encryption and Decryption Network for Internet of Medical Things. *IEEE Internet of Things Journal*, *8*(3), 1504–1518. https://doi.org/10.1109/JIOT.2020.3012452

Hasan, M. K., Islam, S., Sulaiman, R., Khan, S., Hashim, A. H. A., Habib, S., Islam, M., Alyahya, S., Ahmed, M. M., Kamil, S., & Hassan, M. A. (2021). Lightweight Encryption Technique to Enhance Medical Image Security on Internet of Medical Things Applications. *IEEE Access*, *9*, 47731–47742. https://doi.org/10.1109/ACCESS.2021.3061710

Kamal, S. T., Hosny, K. M., Elgindy, T. M., Darwish, M. M., & Fouda, M. M. (2021). A new image encryption algorithm for grey and color medical images. *IEEE Access*, *9*, 37855–37865. https://doi.org/10.1109/ACCESS.2021.3063237

Lin, C. H., Wu, J. X., Chen, P. Y., Lai, H. Y., Li, C. M., Kuo, C. L., & Pai, N. S. (2021). Intelligent Symmetric Cryptography with Chaotic Map and Quantum Based Key Generator for Medical Images Infosecurity. *IEEE Access*, *9*, 118624–118639. https://doi.org/10.1109/ACCESS.2021.3107608

Ravichandran, D., Praveenkumar, P., Rayappan, J. B. B., & Amirtharajan, R. (2017). DNA Chaos Blend to Secure Medical Privacy. *IEEE Transactions on Nanobioscience*, *16*(8), 850–858. https://doi.org/10.1109/TNB.2017.2780881

Shankar, K., Elhoseny, M., Chelvi, E. D., Lakshmanaprabu, S. K., & Wu, W. (2018). An efficient optimal key based chaos function for medical image security. *IEEE Access*, *6*, 77145–77154. https://doi.org/10.1109/ACCESS.2018.2874026

Vaseghi, B., Mobayen, S., Hashemi, S. S., & Fekih, A. (2021). Fast Reaching Finite Time synchronization Approach for Chaotic Systems with Application in Medical Image Encryption. *IEEE Access*, *9*, 25911–25925. https://doi.org/10.1109/ACCESS.2021.3056037

## Authors' Profiles

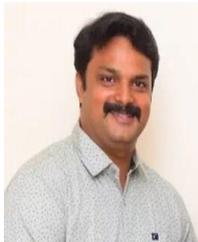

Dr. Chandra Sekhar Sanaboina is an assistant professor currently employed by the JNTUK University Kakinada's University College of Engineering in the Computer Science and Engineering department. From Koneru Lakshmaiah College of Engineering, he graduated with a Bachelor of Technology (B. Tech) in Electronics and Computer Science Engineering in 2005. At Vellore Institute of Technology, he then earned a Master of Technology (M. Tech) in Computer Science and Engineering in 2008. 2020 saw the completion of his Doctor of Philosophy (Ph. D.) at JNTUK in the field of the Internet of Things.

He had been a teacher for more than 12 years and a researcher for about 9 years. His research interests include Artificial Intelligence, Data Science, Cryptography, Machine Learning, the Internet of Things, and wire-free sensor networks.

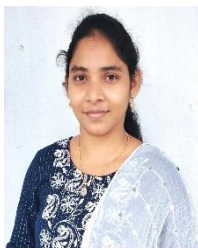

Tejaswini Yadla is a student at the University College of Engineering Kakinada(A), Jawaharlal Nehru Technological University, Kakinada where she is working toward her Masters (M. Tech) in cyber security. She received her B. Tech degree in Computer Science and Engineering (CSE) from AKRG College of Engineering and Technology, Andhra Pradesh in 2016.

Her research interests cover Cryptography, Network Security, Information Security, Digital Forensics, Image Processing, and Optimization Techniques.